\documentclass[conference]{IEEEtran}
\IEEEoverridecommandlockouts
\usepackage{cite}
\usepackage{amsmath,amssymb,amsfonts}
\usepackage{algorithmic}
\usepackage{graphicx}
\usepackage{subcaption}
\usepackage{textcomp}
\usepackage{xcolor}
\usepackage{multirow}
\def\BibTeX{{\rm B\kern-.05em{\sc i\kern-.025em b}\kern-.08em
    T\kern-.1667em\lower.7ex\hbox{E}\kern-.125emX}}
\begin{document}

\title{Improving scDiffusion with Sparsity-Biased Classifier-Free Guidance
%\thanks{This work was supported by JST CREST, Japan, under Grant JPMJCR25T4.}
}

\author{
	\IEEEauthorblockN{
		Yu Song\textsuperscript{1},
		Hao Sun\textsuperscript{1},
		Ikuko Nishikawa\textsuperscript{1},
		and Yen-Wei Chen\textsuperscript{1}
	}
	\IEEEauthorblockA{\textsuperscript{1}
		College of Information Science and Engineering, Ritsumeikan University, Osaka, Japan}
	\thanks{
		Yu Song and Hao Sun contributed equally to this work.
		Corresponding author: Yu Song (yusong@fc.ritsumei.ac.jp) and Yen-Wei Chen (chen@is.ritsumei.ac.jp).
	}
}

\maketitle

\begin{abstract}
Single-cell RNA sequencing (scRNA-seq) has become an essential tool in modern cellular biology, and generating accurate synthetic scRNA-seq data is becoming increasingly important. Although diffusion models have achieved promising results in conditional scRNA-seq generation, existing guidance strategies—including classifier guidance and classifier-free guidance (CFG)—rely on an unconditional branch trained to approximate the true marginal distribution, which may retain substantial gene-specific structure and limit guidance effectiveness. Inspired by recent work showing that diffusion models can be effectively guided using intentionally degraded references, we propose a sparsity-biased classifier-free guidance (SB-CFG) strategy for scRNA-seq generation. Rather than approximating the assumed "neutral" marginal distribution, SB-CFG introduces a deliberately under-informative sparse reference for the unconditional branch, removing gene identity while preserving only coarse sparsity statistics. This “bad” reference amplifies the contrast between conditional and unconditional predictions, leading to stronger and more effective guidance during sampling. We evaluated SB-CFG as a training-free sampling modification on five publicly available scRNA-seq datasets. Experimental results demonstrate consistent improvements over standard CFG-based sampling in terms of marker gene expression fidelity, cell-type consistency, and sparsity preservation, indicating that SB-CFG better captures biologically meaningful gene expression patterns.
\end{abstract}

\begin{IEEEkeywords}
single-cell RNA sequencing, diffusion models, classifier-free guidance, generative modeling
\end{IEEEkeywords}

\section{Introduction}
Single-cell RNA sequencing (scRNA-seq) enables genome-wide gene expression profiling at single-cell resolution, revealing cellular heterogeneity, rare cell populations, and dynamic biological processes that are obscured in bulk measurements \cite{tang2009mrna, kolodziejczyk2015technology}. It has become a foundational technology for cell atlas construction, developmental biology, and disease research, driving the need for robust computational models to analyze single-cell data \cite{luecken2019current, regev2017human}. However, despite technological advances, obtaining sufficient high-quality scRNA-seq data remains challenging due to high cost, experimental complexity, and limited sample availability \cite{suva2019single, stegle2015computational}, motivating the development of generative models that can synthesize realistic single-cell data to support downstream analysis.

Deep learning–based methods have become the dominant approach for  \textit{in silico}  scRNA-seq data generation, following a development trajectory similar to that of image generative models. The most commonly used generative frameworks for scRNA-seq data include variational autoencoders (VAEs) \cite{kingma2013auto}, generative adversarial networks (GANs) \cite{goodfellow2020generative}, and more recently, diffusion models \cite{ho2020denoising, song2020score}. VAE-based methods are effective at learning compact latent representations and have been widely applied to downstream tasks such as batch correction and clustering; however, they often exhibit limited generation quality and face challenges in conditional generation. Representative examples include scVI and scVAE \cite{lopez2018deep, gronbech2020scvae}. GAN-based methods directly model the data distribution and have been applied to single-cell data generation, such as scGAN \cite{marouf2020realistic}. Nevertheless, GANs are known to suffer from unstable training and mode collapse, which limits their robustness and generalizability across datasets \cite{saxena2021generative}.

Diffusion-based generative models have recently achieved strong performance in high-dimensional generation tasks, including image, video, and language modeling \cite{dhariwal2021diffusion, wan2025wan, nie2025large}. From a mathematical perspective, diffusion models define a stochastic or deterministic process that gradually perturbs data toward a simple prior distribution, typically a Gaussian distribution. By learning the reverse process, the model can iteratively denoise samples from the prior back to the data space using numerical solvers such as Euler or Heun methods. Existing diffusion models can be categorized based on their learning targets, including noise prediction, score function estimation, flow matching, or direct data prediction with reparameterized losses \cite{song2020score, lipman2022flow, liu2022flow, li2025back}. Furthermore, diffusion can be performed either in the original data space or in a learned latent space, where latent diffusion models leverage an autoencoder to reduce dimensionality and improve efficiency \cite{rombach2022high}.

Building on these advances, scDiffusion introduced latent diffusion modeling to the scRNA-seq domain \cite{luo2024scdiffusion}. In scDiffusion, a pretrained foundation model based on a variational autoencoder architecture, SCimilarity, is first used to project high-dimensional gene expression profiles into a low-dimensional latent space \cite{heimberg2025cell}. A diffusion model is then trained in this latent space to model the data distribution, while a separate classifier is trained to enable conditional generation. During sampling, classifier guidance is applied by adding the gradient of the classifier with respect to the latent variable to the diffusion model’s score function, thereby steering generation toward specific cell types or biological conditions \cite{dhariwal2021diffusion}.

While effective, classifier-guided diffusion models introduce several limitations. First, training an auxiliary classifier increases model complexity and computational cost. Second, classifier guidance relies on accurate classification across all diffusion timesteps, which can be unreliable under high noise levels. More importantly, both classifier guidance and standard classifier-free guidance (CFG) implicitly assume that the unconditional branch provides a neutral reference distribution \cite{ho2022classifier}. For highly sparse scRNA-seq data, this assumption is problematic, as the unconditional prediction may still retain gene-specific structure and thus weaken the guidance contrast during sampling.

Motivated by these observations and recent work demonstrating that diffusion models can be effectively guided using intentionally degraded references \cite{karras2024guiding}, we propose a sparsity-biased classifier-free guidance (SB-CFG) strategy as a training-free sampling modification. Instead of using the model’s learned unconditional prediction directly, SB-CFG replaces it with an intentionally under-informed sparse baseline that preserves only global sparsity statistics while discarding gene identity information. By constructing a “worse” unconditional reference during sampling, SB-CFG amplifies the contrast between conditional and unconditional predictions, leading to stronger and more effective guidance toward conditional gene expression patterns.

In summary, the main contributions of this work are as follows:
\begin{itemize}
	\item We analyze the limitations of existing classifier-guided and CFG-based diffusion sampling strategies for sparse scRNA-seq data;
	\item We propose a biologically motivated, training-free sparsity-biased CFG (SB-CFG) sampling mechanism inspired by the principle of guiding diffusion with a degraded reference;
	\item We demonstrate that the proposed SB-CFG consistently improves conditional generation quality and biological fidelity across multiple public scRNA-seq datasets.
\end{itemize}

\section{Proposed Method}

\begin{figure*}[htbp]
	\centering
	\includegraphics[width=0.66\textwidth]{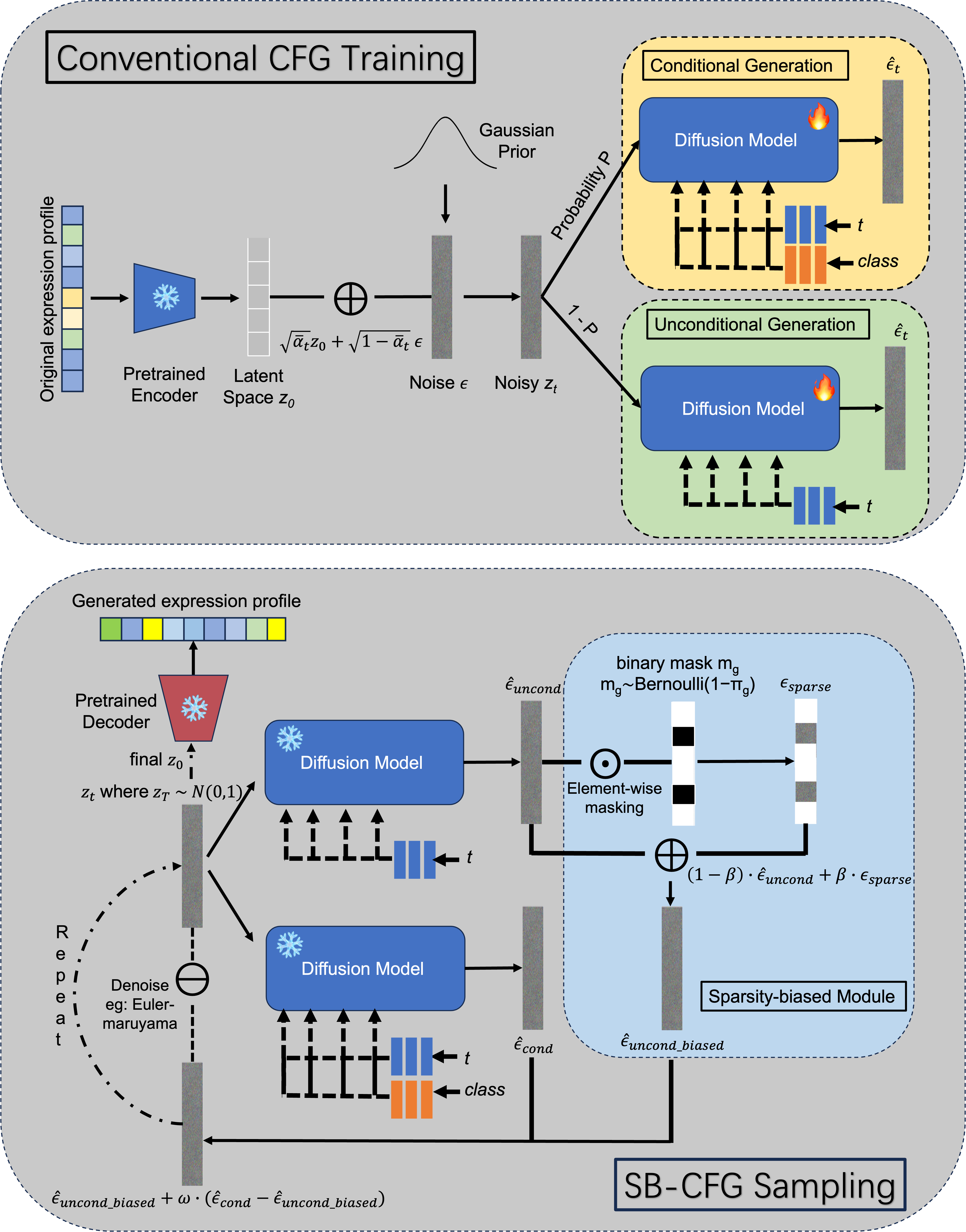}
	\caption{Overview of the proposed method. During training, we follow the standard scDiffusion framework and adopt classifier-free guidance (CFG) without any architectural modification. During sampling, we replace conventional CFG with the proposed sparsity-biased classifier-free guidance (SB-CFG). When the sparsity-biased module is removed, the sampling procedure reduces to standard CFG, highlighting SB-CFG as a drop-in, training-free modification. }
	\label{fig:overview}
\end{figure*}

In this research, we propose \textbf{Sparsity-Biased Classifier-Free Guidance (SB-CFG)}, a training-free sampling modification for conditional diffusion models tailored to single-cell RNA sequencing (scRNA-seq) data, as shown in Figure~\ref{fig:overview}. SB-CFG is designed to address the mismatch between existing diffusion guidance mechanisms and the intrinsic sparsity structure of scRNA-seq data.

Our method builds upon the scDiffusion framework, a latent diffusion model that combines a pretrained VAE from SCimilarity \cite{heimberg2025cell} with a diffusion model operating in a one-dimensional latent space to perform conditional generation. The original scDiffusion relies on classifier guidance, which requires training separate classifiers for each condition. Following Zhang et al.~\cite{zhang2025cfdiffusion}, we first replaced classifier guidance with CFG to enable more flexible conditional generation without additional classifier training. 
Building on this CFG-based framework, we then introduce SB-CFG, which intervenes only at the sampling stage by replacing the unconditional reference with an intentionally degraded, sparsity-preserving gradient. This gradient removes gene-specific information while preserving global sparsity characteristics, thereby enhancing guidance contrast and improving conditional generation quality for sparse scRNA-seq data.

\subsection{Assumption of a 'neutral' prior in scRNA-seq data}

CFG treats the unconditional model output 
$\epsilon_{\text{uncond}}$ as representing a neutral prior distribution $p(z)$. 
Specifically, CFG assumes that
\begin{equation}
	\epsilon_{\text{uncond}} \approx -\sigma_t \nabla_z \log p(z_t),
\end{equation}
where $\sigma_t$ is the noise standard deviation at timestep t.
This unconditional output is intended to encode no information about the condition $y$. Under this assumption, conditional information is introduced by amplifying the difference between conditional and unconditional predictions during sampling.

However, for real scRNA-seq data, this neutrality assumption does not hold.
The prior $p(z)$ learned by the unconditional model corresponds to the empirical
marginal distribution aggregated over all cell types.
Unlike image data, scRNA-seq data are highly structured and extremely sparse,
with typically more than 80--90\% zero entries in the gene expression matrix.
Each gene follows a highly non-uniform marginal distribution, often modeled
using zero-inflated distributions.
As a result, the unconditional diffusion model inevitably learns strong
gene-specific baselines, such as pushing most genes toward zero expression or
toward their global mean.
Consequently, $\epsilon_{\text{uncond}}$ reflects gene-level biases inherited
from the marginal $p(z)$, rather than a flat or neutral prior.

This effect reduces the effective contrast used for guidance.
The CFG update relies on the difference
$\epsilon_{\text{cond}} - \epsilon_{\text{uncond}}$.
For a gene $i$ that is almost always zero across cell types, all conditions tend to agree on its behavior (e.g., all favor low expression). Consequently, $\nabla_i \log p(z \mid y)$ will be similar across all $y$, and the marginal score $\nabla_i \log p(z)$, being a weighted average, closely matches each conditional score.
As a result, the guided prediction
\begin{equation}
	\tilde{\epsilon}
	=
	\epsilon_{\text{uncond}}
	+
	w \bigl(\epsilon_{\text{cond}} - \epsilon_{\text{uncond}}\bigr)
\end{equation}
contains only a weak contribution along such gene dimensions.

Because the unconditional model already encodes gene-level sparsity patterns from the marginal distribution, subtracting it can inadvertently suppress condition-specific expression signals. This violates CFG's underlying assumption of a neutral, condition-agnostic prior, limiting its effectiveness for highly sparse scRNA-seq data.

\subsection{Sparsity-Biased Classifier-Free Guidance (SB-CFG)}

To address the issue inherent in standard CFG mentioned above, we introduce \emph{Sparsity-Biased Classifier-Free Guidance} (SB-CFG), a sampling-only modification designed for highly sparse data such as scRNA-seq. SB-CFG mitigates the neutrality assumption issue by constructing a deliberately under-informative, sparsity-preserving baseline, enhancing the effectiveness of conditional guidance.

We first construct a sparse version of the unconditional noise prediction $\epsilon_{\text{sparse}}$ by applying a stochastic binary mask to $\epsilon_{\text{uncond}}$, as illustrated in Figure~\ref{fig:overview}. For each dimension $i$, we sample a Bernoulli mask that zeros out the prediction with probability $\pi_i$:
\begin{equation}
	\epsilon_{\text{sparse}} = m \odot \epsilon_{\text{uncond}}, \quad m_i \sim \text{Bernoulli}(1 - \pi_i),
\end{equation}
where $\pi_i$ is the empirical probability that gene $i$ has zero expression in the dataset. Intuitively, setting $\epsilon_{\text{uncond},i}$ to zero with probability $\pi_i$ is a neutral operation: it does not introduce gene-specific directional bias, but only enforces the expected sparsity level observed in real data.

To bias the unconditional branch in a controlled manner, we blend the original prediction $\epsilon_{\text{uncond}}$ with its sparse version using a mixing coefficient $\beta \in [0, 1]$:
\begin{equation}
	\epsilon_{\text{uncond\_bias}} = (1 - \beta)\,\epsilon_{\text{uncond}} + \beta\,\epsilon_{\text{sparse}}.
\end{equation}
This interpolation yields a sparsity-biased unconditional prediction that preserves the expected sparsity level while reducing gene-specific information. When $\beta = 0$, SB-CFG reduces to standard CFG; when $\beta = 1$, the unconditional prediction is fully replaced by the sparse version.

Finally, we apply CFG-style guidance using the biased unconditional branch. Given a conditional prediction $\epsilon_{\text{cond}}$ and guidance weight $w$, the final guided prediction is:
\begin{equation}
	\tilde{\epsilon}_{\text{SB}} = \epsilon_{\text{uncond\_bias}} + w\,\left(\epsilon_{\text{cond}} - \epsilon_{\text{uncond\_bias}}\right).
\end{equation}

The key advantage of SB-CFG lies in restoring a neutral baseline by deliberately weakening the unconditional reference. Since $\epsilon_{\text{sparse}}$ lacks gene-specific content but maintains realistic sparsity, the difference $\epsilon_{\text{cond}} - \epsilon_{\text{uncond\_bias}}$ captures condition-specific signals more effectively. SB-CFG thus improves conditional sampling in sparse data regimes without requiring retraining, offering a principled enhancement over standard CFG.

\begin{table*}[!t]
	\centering
	\caption{Quantitative comparison of SB-CFG and standard CFG on five public datasets. Both used identical trained diffusion weights and differed only in sampling.}
	\label{tab_result}
	\begin{tabular}{l cc cc cc cc}
		\hline
		\multirow{2}{*}{Dataset} 
		& \multicolumn{2}{c}{Pearson $r \, \uparrow$} 
		& \multicolumn{2}{c}{Cell-Type Cls $\uparrow$} 
		& \multicolumn{2}{c}{Zero-Rate Diff $\downarrow$} 
		& \multicolumn{2}{c}{Marker Specificity $\uparrow$} \\
		\cline{2-9}
		& CFG & SB-CFG & CFG & SB-CFG & CFG & SB-CFG & CFG & SB-CFG \\
		\hline
		  Baron Human \cite{baron2016single}      
		& 0.81 & \textbf{0.91} 
		& 0.27 & \textbf{0.73} 
		& 0.58 & \textbf{0.57} 
		& 1.22 & \textbf{2.50} \\
		
		Baron Mouse \cite{baron2016single}      
		& 0.89 & \textbf{0.97} 
		& 0.28 & \textbf{0.62} 
		& 0.54 & \textbf{0.53} 
		& 1.34 & \textbf{1.95} \\
		
		Human Lung \cite{habermann2020single}        
		& 0.94 & \textbf{0.95} 
		& 0.34 & \textbf{0.89} 
		& 0.48 & \textbf{0.45} 
		& 1.94 & \textbf{2.67} \\
		
		Mizrak  \cite{mizrak2019single} 
		& 0.89 & \textbf{0.94} 
		& 0.23 & \textbf{0.86} 
		& \textbf{0.56} & \textbf{0.56} 
		& 1.04 & \textbf{2.24} \\
		
		PBMC68k \cite{zheng2017massively}           
		& 0.93 & \textbf{0.95} 
		& 0.24 & \textbf{0.63} 
		& \textbf{0.60} & 0.61 
		& 0.71 & \textbf{1.46} \\
		\hline
	\end{tabular}
\end{table*}

\begin{figure*}[!t]
	\centering
	
	% Row 1
	\begin{subfigure}[t]{0.385\textwidth}
		\centering
		\includegraphics[width=\textwidth]{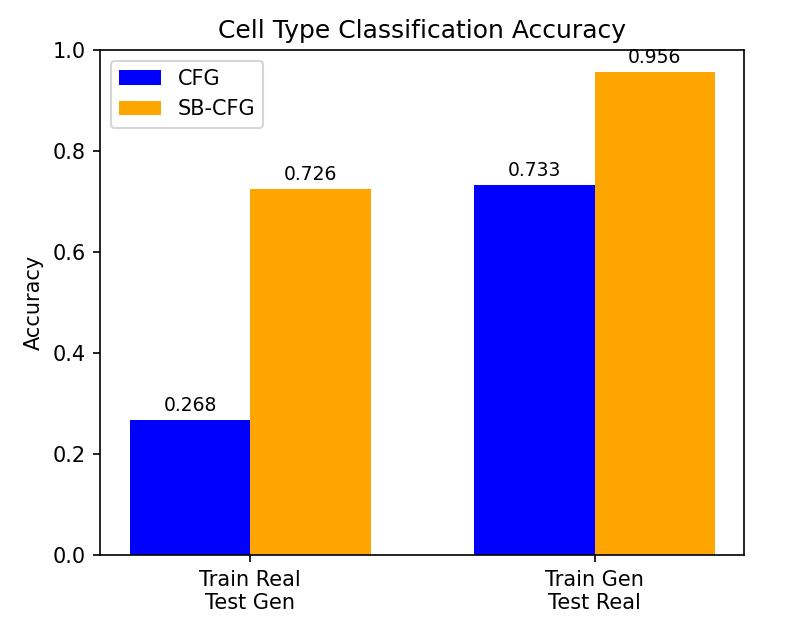}
		\caption{CFG vs. SB-CFG classification accuracy. Left: train on real data and test on generated data. Right: train on generated data and test on real data.}
		\label{fig:sub1}
	\end{subfigure}
	\hfill
	\begin{subfigure}[t]{0.29\textwidth}
		\centering
		\includegraphics[width=\textwidth]{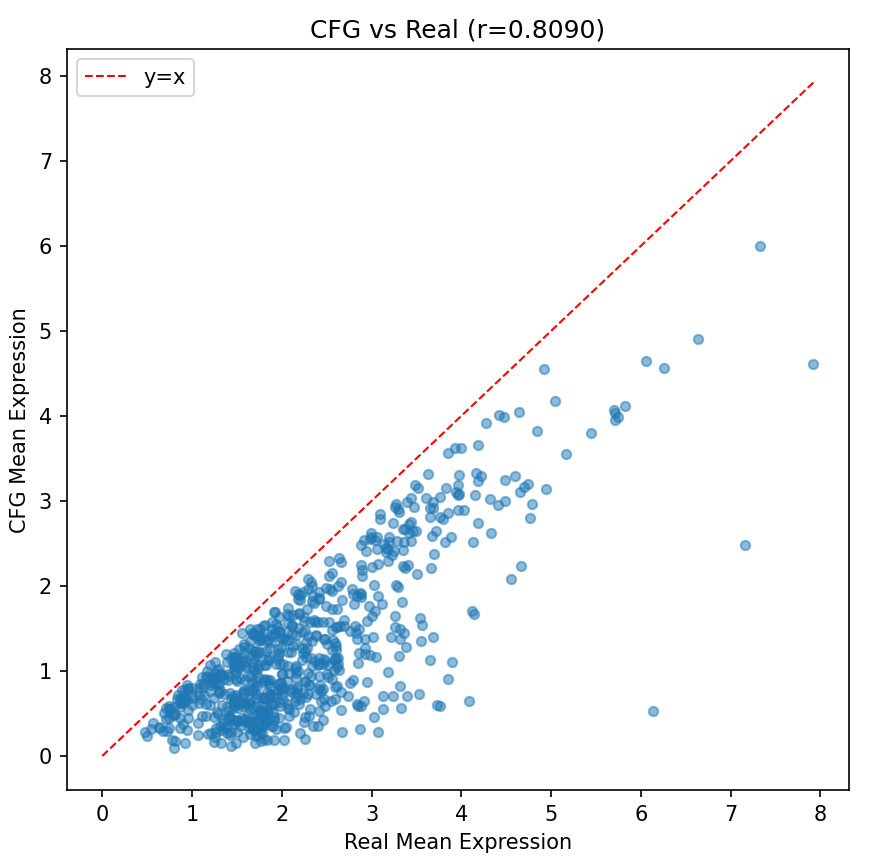}
		\caption{CFG Pearson correlation between the mean gene expression profiles of real and generated cells.}
		\label{fig:sub2}
	\end{subfigure}
	\hfill
	\begin{subfigure}[t]{0.29\textwidth}
		\centering
		\includegraphics[width=\textwidth]{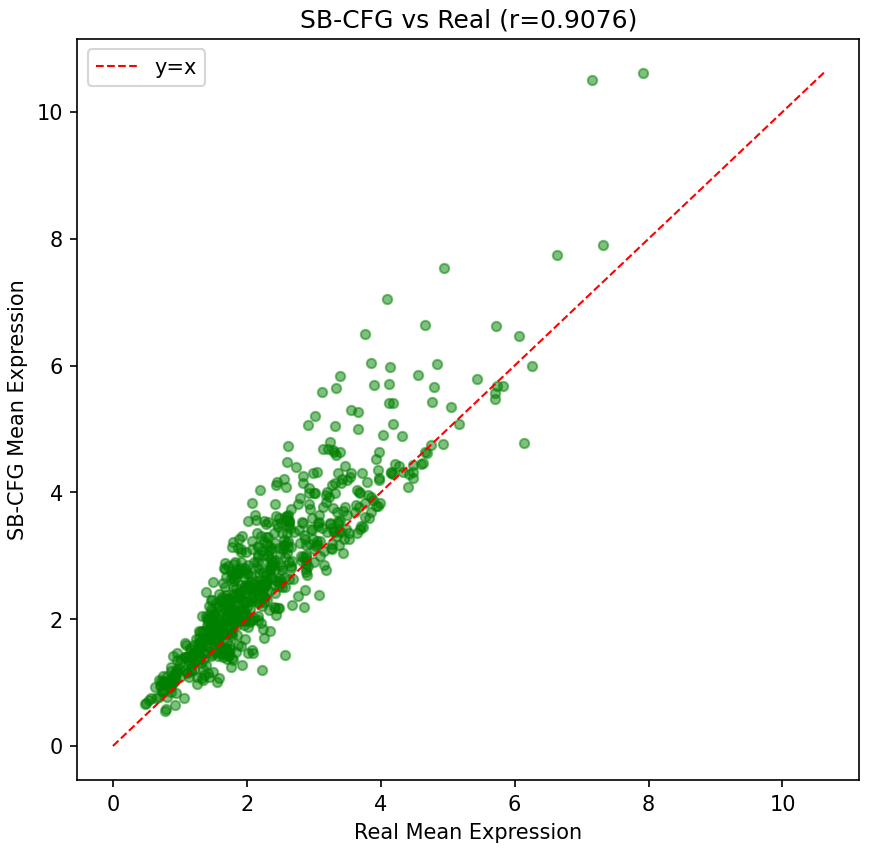}
		\caption{SB-CFG Pearson correlation between the mean gene expression profiles of real and generated cells.}
		\label{fig:sub3}
	\end{subfigure}
	
	\vspace{0.5em}
	
	% Row 2
	\begin{subfigure}[t]{0.22\textwidth}
		\centering
		\includegraphics[width=\textwidth]{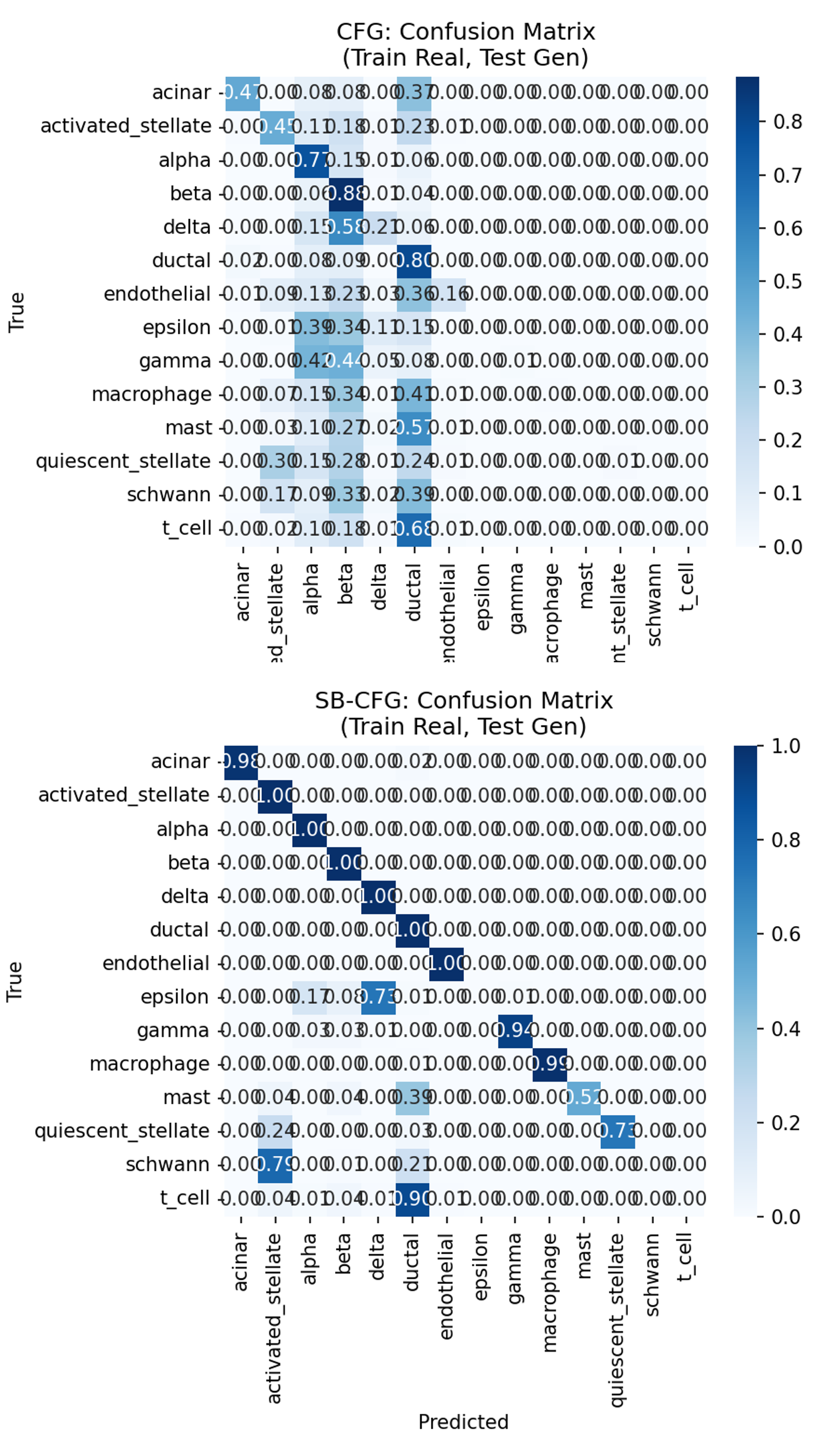}
		\caption{Per-class confusion matrix for CFG vs. SB-CFG. Rows indicate predicted classes, and columns indicate true (real) classes.}
		\label{fig:sub4}
	\end{subfigure}
	\hfill
	\begin{subfigure}[t]{0.30\textwidth}
		\centering
		\includegraphics[width=\textwidth]{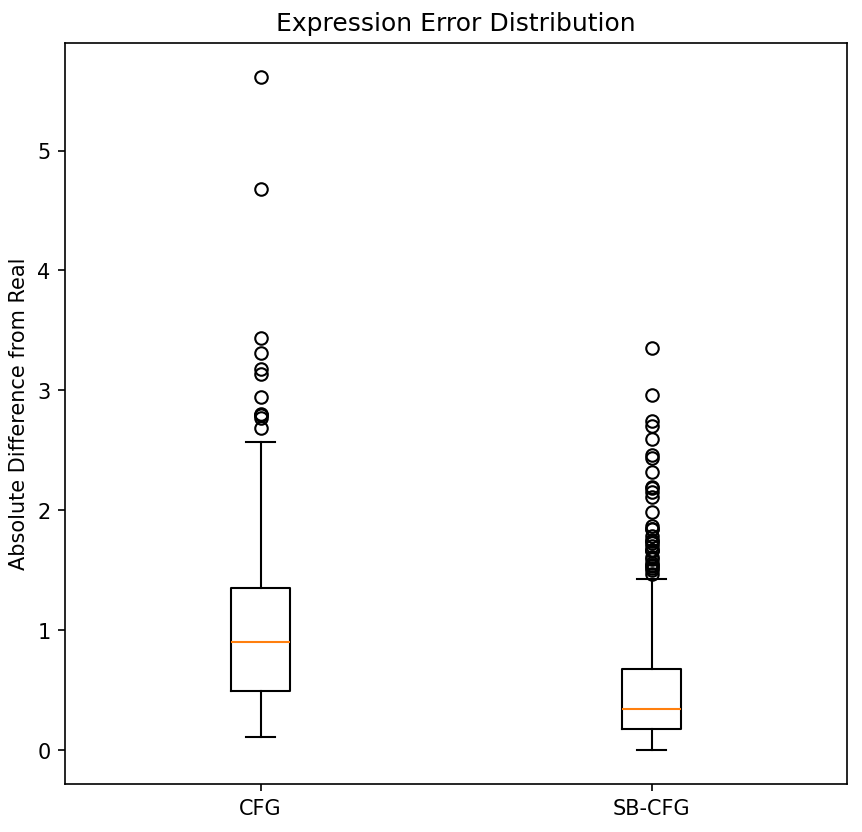}
		\caption{CFG vs. SB-CFG absolute expression error between real and generated cell profiles.}
		\label{fig:sub5}
	\end{subfigure}
	\hfill
	\begin{subfigure}[t]{0.30\textwidth}
		\centering
		\includegraphics[width=\textwidth]{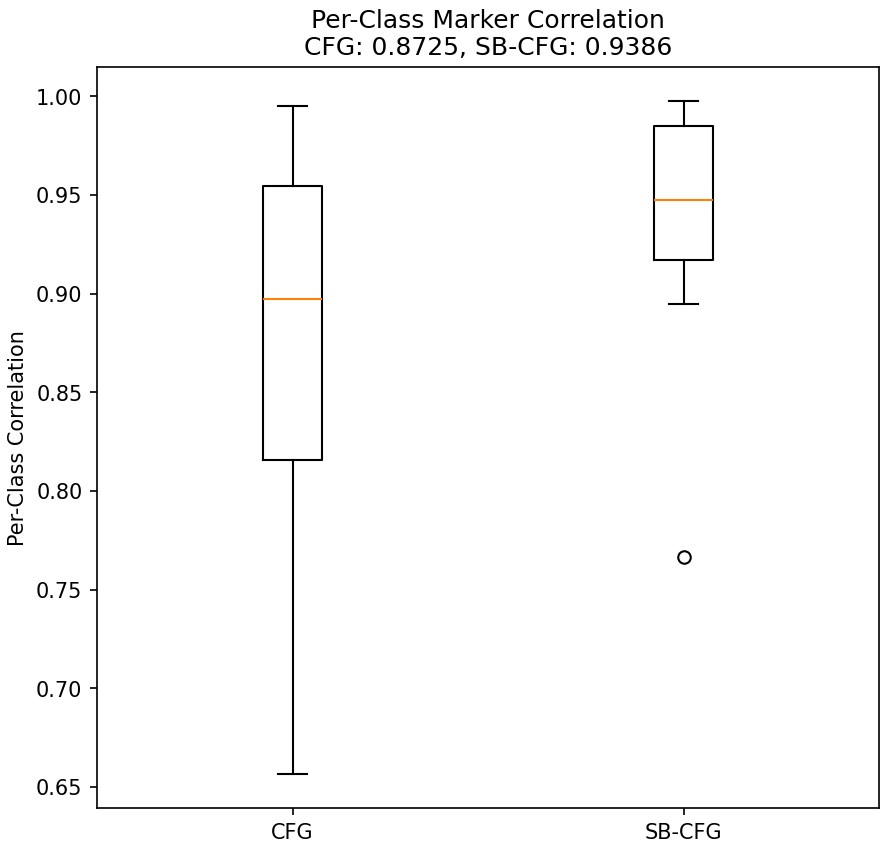}
		\caption{CFG vs. SB-CFG per-class marker gene correlation between real and generated cell profiles.}
		\label{fig:sub6}
	\end{subfigure}
	
	\caption{Visualization of evaluation metrics on the Baron Human dataset.}
	\label{fig:main_6subfig}
\end{figure*}

\begin{figure*}[t]
	\centering
	\includegraphics[width=0.95\textwidth]{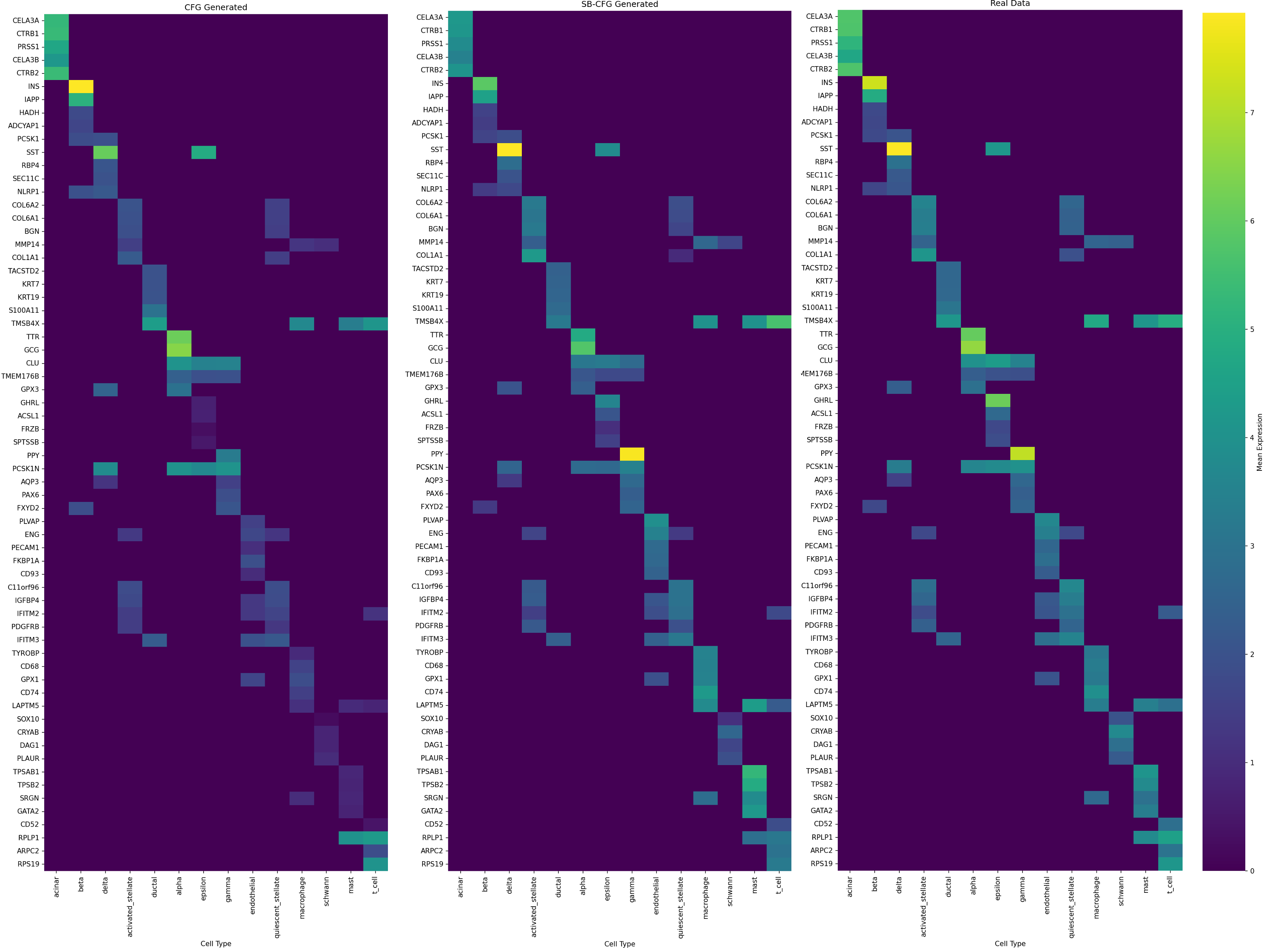}
	\caption{Marker gene heatmaps for different cell classes in the Baron Human dataset. Left: CFG, middle: SB-CFG, right: Real.}
	\label{fig:heatmap}
\end{figure*}

\begin{table*}[t]
\caption{Ablation on the mixing coefficient $\beta$ across five datasets. 
Each cell shows Pearson $r\uparrow$ / Cell-Type Cls Acc $\uparrow$ / 
Zero-Rate Diff $\downarrow$ / Marker Specificity $\uparrow$. 
Bold indicates the best $\beta$ per metric per dataset (ties bolded). 
$\beta=0$ corresponds to standard CFG; $\beta=0.5$ is the default used in our main experiments (Table~\ref{tab_result}).}
\label{tab:beta_ablation}
\centering
\footnotesize
\begin{tabular}{c|ccccc}
\hline
$\beta$ & Baron Human & Baron Mouse & Human Lung & Mizrak & PBMC68k \\
\hline
0.00 & 0.81 / 0.27 / 0.58 / 1.22 & 0.89 / 0.28 / 0.54 / 1.34 & 0.94 / 0.34 / 0.48 / 1.94 & 0.89 / 0.23 / 0.56 / 1.04 & 0.93 / 0.24 / \textbf{0.60} / 0.71 \\
0.25 & 0.87 / 0.73 / \textbf{0.49} / \textbf{2.53} & \textbf{0.97} / 0.57 / \textbf{0.50} / 1.93 & 0.91 / 0.84 / 0.45 / \textbf{2.68} & \textbf{0.94} / 0.85 / \textbf{0.46} / 2.22 & \textbf{0.95} / \textbf{0.67} / 0.62 / 1.45 \\
0.50 & \textbf{0.91} / 0.73 / 0.57 / 2.50 & \textbf{0.97} / \textbf{0.62} / 0.53 / 1.95 & 0.95 / \textbf{0.89} / 0.45 / 2.67 & \textbf{0.94} / \textbf{0.86} / 0.56 / \textbf{2.24} & \textbf{0.95} / 0.63 / 0.61 / \textbf{1.46} \\
0.75 & \textbf{0.91} / \textbf{0.79} / 0.53 / 2.38 & \textbf{0.97} / 0.59 / \textbf{0.50} / \textbf{1.96} & 0.96 / 0.86 / \textbf{0.44} / 2.64 & 0.93 / 0.85 / 0.50 / 2.13 & 0.94 / 0.57 / 0.62 / 1.42 \\
1.00 & 0.90 / 0.71 / 0.54 / 2.18 & 0.96 / 0.60 / 0.51 / 1.95 & \textbf{0.97} / 0.88 / \textbf{0.44} / 2.62 & 0.93 / \textbf{0.86} / 0.52 / 1.95 & \textbf{0.95} / 0.51 / 0.63 / 1.33 \\
\hline
\end{tabular}
\end{table*}

\section{Experiments}
\subsection{Dataset}
We evaluated our model on five commonly used scRNA-seq datasets from both human and mouse.
The Baron Human dataset \cite{baron2016single} contains single-cell transcriptomes of human pancreatic islets from four donors, profiling over 8,000 cells across multiple endocrine and non-endocrine cell types.
The Baron Mouse dataset \cite{baron2016single} similarly profiles pancreatic cells from two mouse strains using the same droplet-based scRNA-seq approach.
The Human Lung Pulmonary Fibrosis (PF) dataset \cite{habermann2020single} includes more than 110,000 human lung cells from pulmonary fibrosis samples.
The Mizrak dataset \cite{mizrak2019single} contains 6,022 mouse spleen cells with 16 cell types, including T cell subsets (CD4 T, CD8 T, Treg, NK), B cell subpopulations, dendritic cells, and macrophages.
The PBMC68k dataset \cite{zheng2017massively} contains human peripheral blood mononuclear cells (PBMCs) with 11 immune cell types. Following scDiffusion, we excluded CD4$^+$ T helper 2 cells.
For all datasets, we removed cells with fewer than 10 total counts and removed genes expressed in fewer than 3 cells.

\subsection{Implementation Details}
The diffusion model used $T = 1000$ timesteps with a linear noise schedule, and operated in a $128$-dimensional VAE latent space. 
The VAE (SCimilarity\cite{heimberg2025cell}) consisted of an encoder with three hidden layers of $1024$ units each, which maps gene expression vectors (e.g., $27{,}281$ genes for the Human Lung dataset) into the $128$-dimensional latent space, and a symmetric decoder for reconstruction. 
The diffusion network was a U-Net style MLP with hidden dimensions $[512, 512, 256, 128]$, trained with classifier-free guidance using a class dropout probability of $0.2$.

For SB-CFG sampling, we used a sparsity mixing weight $\beta=0.5$ (selected via the ablation study in Section \ref{sec:ablation}) to blend the original unconditional prediction with its sparse version, and set the guidance scale to $\omega = 3.0$ to control the strength of class conditioning.
The sparsity bias was applied only during the final 30\% of denoising timesteps (when $t/T < 0.3$), selected from \{10\%, 30\%, 50\%, 100\%\}
via ablation (full results omitted due to space constraints). Gene-wise zero probabilities $\pi$ were computed from the training data using a threshold of 0.01 on log1p-transformed expression values. This threshold corresponds to raw counts that are effectively zero after normalization, consistent with standard scRNA-seq preprocessing practice, and we verified empirically that values in the range [0.001,0.1] yielded comparable sparsity estimates.
A new random mask was sampled at every diffusion timestep, introducing stochasticity that helps explore sparse solutions. 
When the VAE latent dimension differed from the number of genes, we used the mean zero probability across all genes as a fallback. 

Importantly, standard CFG and SB-CFG sampling shared the same pre-trained CFG diffusion model weights, and no additional training was required.

\subsection{Evaluation Metrics}

\paragraph{Pearson Correlation Coefficient($r \uparrow$)}
We computed the Pearson Correlation Coefficient $r$ between the mean gene expression profiles of real and generated cells.
The mean profiles were computed for each cell type and then aggregated across all cell types.
This metric measures how well the generated data keeps the overall gene expression patterns.
Higher values indicate better agreement with real data.

\paragraph{Cell-Type Classification Accuracy ($\uparrow$)}
We trained a Random Forest classifier with 100 trees on real cells and tested it on generated cells.
This metric checks whether generated cells show clear cell-type expression patterns.
Higher accuracy means the generated cells are easier to classify into the correct cell types.

\paragraph{Zero-Rate Difference ($\downarrow$)}
As mentioned, scRNA-seq data has many zeros.
We computed the absolute difference in zero rates between real and generated data, averaged across genes.
Lower values mean the generated data better matches the sparsity level of real data.

\paragraph{Marker Gene Specificity ($\uparrow$)}
Marker genes are expected to have high expression in their target cell type and low expression in other cell types.
For each marker gene, we computed a specificity score as the ratio between its mean expression in the target cell type and its mean expression in all other cell types.
We compared these scores between generated and real data.
Scores closer to the real data indicate better preservation of cell-type marker patterns.

\subsection{Experimental Results}

For each dataset, we generated 1,000 synthetic cells per cell type using both standard CFG and SB-CFG. Both methods used identical pre-trained VAE and diffusion model weights, since SB-CFG is purely a training-free modification applied during sampling. The quantitative results are summarized in Table \ref{tab_result}. Figure \ref{fig:main_6subfig} presents qualitative comparisons on the Baron Human dataset. SB-CFG outperformed CFG across all visualization metrics, including classification accuracy, mean expression correlation, per-class confusion, expression error distribution, and per-class marker gene correlation. Figure \ref{fig:heatmap} further shows marker gene heatmaps across cell types in the Baron Human dataset, where SB-CFG-generated samples more closely match the real expression profiles than those from standard CFG.

\subsection{Ablation Study}
\label{sec:ablation}
To investigate the sensitivity of SB-CFG to the mixing coefficient $\beta$, we conducted an ablation study across all five datasets by varying $\beta \in \{0.0, 0.25, 0.5, 0.75, 1.0\}$ while keeping all other settings fixed. $\beta=0.0$ corresponds to standard CFG (no sparsity bias), while $\beta=1.0$ fully replaces the unconditional prediction with its sparse version. Results are summarized in Table~\ref{tab:beta_ablation}.

The ablation reveals two consistent trends. First, introducing even moderate sparsity bias ($\beta=0.25$) substantially improves Pearson correlation across all datasets, confirming that the unconditional prediction in standard CFG retains gene-specific structure that hinders guidance. Second, there is a trade-off between global expression fidelity and cell-type discrimination: increasing 
$\beta$ generally improves Pearson $r$ but can reduce classification accuracy at high values, particularly on datasets with many fine-grained cell types such as PBMC68k. $\beta=0.5$ provides a robust balance across both objectives on all five datasets, motivating its use as the default in our main experiments. The dataset-dependent variation in the optimal $\beta$ further supports our identification of adaptive $\beta$ selection as a promising direction for future work.

\section{Conclusion}
In this paper, we proposed Sparsity-Biased Classifier-Free Guidance (SB-CFG), a simple yet effective sampling modification for conditional generation of scRNA-seq data using diffusion models. The core contribution is the biological insight that the extreme sparsity of scRNA-seq data structurally violates the neutrality assumption underlying standard CFG: the unconditional model already 
encodes gene-level sparsity patterns from the marginal distribution, weakening the guidance signal for condition-specific genes. SB-CFG addresses this by replacing the unconditional output with a sparse, less informative baseline, amplifying the contrast between conditional and unconditional predictions in a principled way. The method requires no additional training or architectural changes, 
and only needs gene-level zero-expression probabilities computed from the training data. Experiments across five diverse scRNA-seq datasets demonstrate consistent improvements over standard CFG in both statistical quality and biological relevance.

Despite these improvements, several limitations remain. 
First, our formulation uses a single global zero-probability 
vector $\pi$, assuming sparsity is shared across conditions; 
datasets with cell-type-specific dropout could benefit from 
condition-specific $\pi_y$. Second, while $\beta=0.5$ is a 
robust default, the optimal value varies across datasets, 
suggesting adaptive or data-driven strategies could yield 
further gains. Third, a broader comparison against alternative 
guidance strategies, such as the autoguidance of Karras 
et al.~\cite{karras2024guiding}, would further situate SB-CFG 
in the landscape of guidance methods. Future work could also 
extend SB-CFG to other sparse modalities such as single-cell 
ATAC-seq and spatial transcriptomics.

\section*{Ethics Statement}
No new experimental procedures involving human subjects or animals were performed in this study. All analyses were conducted on previously published, publicly available scRNA-seq datasets~\cite{baron2016single, habermann2020single, mizrak2019single, zheng2017massively}. for which the original experimental procedures involving human subjects and animal models were approved by the Institutional Review Boards and Institutional Animal Care and Ethics Committees of the respective contributing institutions, as described in the primary publications.

\section*{Acknowledgment}
This work was supported by JST CREST, Japan, under Grant 
JPMJCR25T4. We would like to thank Prof.~Keiji Nakajima and 
Prof.~Tatsuaki Goh from the Nara Institute of Science and 
Technology for their valuable support and discussions 
throughout this research.


\begin{thebibliography}{00}
\bibitem{tang2009mrna}Tang, F., Barbacioru, C., Wang, Y., Nordman, E., Lee, C., Xu, N., Wang, X., Bodeau, J., Tuch, B., Siddiqui, A. \& Others mRNA-Seq whole-transcriptome analysis of a single cell. {\em Nature Methods}. \textbf{6}, 377-382 (2009)

\bibitem{kolodziejczyk2015technology}Kolodziejczyk, A., Kim, J., Svensson, V., Marioni, J. \& Teichmann, S. The technology and biology of single-cell RNA sequencing. {\em Molecular Cell}. \textbf{58}, 610-620 (2015)

\bibitem{luecken2019current}Luecken, M. \& Theis, F. Current best practices in single-cell RNA-seq analysis: a tutorial. {\em Molecular Systems Biology}. \textbf{15}, e8746 (2019)

\bibitem{regev2017human}Regev, A., Teichmann, S., Lander, E., Amit, I., Benoist, C., Birney, E., Bodenmiller, B., Campbell, P., Carninci, P., Clatworthy, M. \& Others The human cell atlas. {\em Elife}. \textbf{6} pp. e27041 (2017)

\bibitem{suva2019single}Suvà, M. \& Tirosh, I. Single-cell RNA sequencing in cancer: lessons learned and emerging challenges. {\em Molecular Cell}. \textbf{75}, 7-12 (2019)

\bibitem{stegle2015computational}Stegle, O., Teichmann, S. \& Marioni, J. Computational and analytical challenges in single-cell transcriptomics. {\em Nature Reviews Genetics}. \textbf{16}, 133-145 (2015)

\bibitem{kingma2013auto}Kingma, D. \& Welling, M. Auto-encoding variational bayes. {\em ArXiv Preprint ArXiv:1312.6114}. (2013)

\bibitem{goodfellow2020generative}Goodfellow, I., Pouget-Abadie, J., Mirza, M., Xu, B., Warde-Farley, D., Ozair, S., Courville, A. \& Bengio, Y. Generative adversarial networks. {\em Communications Of The ACM}. \textbf{63}, 139-144 (2020)

\bibitem{ho2020denoising}Ho, J., Jain, A. \& Abbeel, P. Denoising diffusion probabilistic models. {\em Advances In Neural Information Processing Systems}. \textbf{33} pp. 6840-6851 (2020)

\bibitem{song2020score}Song, Y., Sohl-Dickstein, J., Kingma, D., Kumar, A., Ermon, S. \& Poole, B. Score-based generative modeling through stochastic differential equations. {\em ArXiv Preprint ArXiv:2011.13456}. (2020)


\bibitem{lopez2018deep}Lopez, R., Regier, J., Cole, M., Jordan, M. \& Yosef, N. Deep generative modeling for single-cell transcriptomics. {\em Nature Methods}. \textbf{15}, 1053-1058 (2018)

\bibitem{gronbech2020scvae}Grønbech, C., Vording, M., Timshel, P., Sønderby, C., Pers, T. \& Winther, O. scVAE: variational auto-encoders for single-cell gene expression data. {\em Bioinformatics}. \textbf{36}, 4415-4422 (2020)

\bibitem{marouf2020realistic}Marouf, M., Machart, P., Bansal, V., Kilian, C., Magruder, D., Krebs, C. \& Bonn, S. Realistic in silico generation and augmentation of single-cell RNA-seq data using generative adversarial networks. {\em Nature Communications}. \textbf{11}, 166 (2020)

\bibitem{saxena2021generative}Saxena, D. \& Cao, J. Generative adversarial networks (GANs) challenges, solutions, and future directions. {\em ACM Computing Surveys (CSUR)}. \textbf{54}, 1-42 (2021)

\bibitem{dhariwal2021diffusion}Dhariwal, P. \& Nichol, A. Diffusion models beat gans on image synthesis. {\em Advances In Neural Information Processing Systems}. \textbf{34} pp. 8780-8794 (2021)

\bibitem{rombach2022high}Rombach, R., Blattmann, A., Lorenz, D., Esser, P. \& Ommer, B. High-resolution image synthesis with latent diffusion models. {\em Proceedings Of The IEEE/CVF Conference On Computer Vision And Pattern Recognition}. pp. 10684-10695 (2022)

\bibitem{wan2025wan}Wan, T., Wang, A., Ai, B., Wen, B., Mao, C., Xie, C., Chen, D., Yu, F., Zhao, H., Yang, J. \& Others Wan: Open and advanced large-scale video generative models. {\em ArXiv Preprint ArXiv:2503.20314}. (2025)

\bibitem{nie2025large}Nie, S., Zhu, F., You, Z., Zhang, X., Ou, J., Hu, J., Zhou, J., Lin, Y., Wen, J. \& Li, C. Large language diffusion models. {\em ArXiv Preprint ArXiv:2502.09992}. (2025)

\bibitem{lipman2022flow}Lipman, Y., Chen, R., Ben-Hamu, H., Nickel, M. \& Le, M. Flow matching for generative modeling. {\em ArXiv Preprint ArXiv:2210.02747}. (2022)

\bibitem{liu2022flow}Liu, X., Gong, C. \& Liu, Q. Flow straight and fast: Learning to generate and transfer data with rectified flow. {\em ArXiv Preprint ArXiv:2209.03003}. (2022)

\bibitem{li2025back}Li, T. \& He, K. Back to basics: Let denoising generative models denoise. {\em ArXiv Preprint ArXiv:2511.13720}. (2025)

\bibitem{luo2024scdiffusion}Luo, E., Hao, M., Wei, L. \& Zhang, X. scDiffusion: conditional generation of high-quality single-cell data using diffusion model. {\em Bioinformatics}. \textbf{40}, btae518 (2024)

\bibitem{heimberg2025cell}Heimberg, G., Kuo, T., DePianto, D., Salem, O., Heigl, T., Diamant, N., Scalia, G., Biancalani, T., Turley, S., Rock, J. \& Others A cell atlas foundation model for scalable search of similar human cells. {\em Nature}. \textbf{638}, 1085-1094 (2025)

\bibitem{ho2022classifier}Ho, J. \& Salimans, T. Classifier-free diffusion guidance. {\em ArXiv Preprint ArXiv:2207.12598}. (2022)

\bibitem{karras2024guiding}Karras, T., Aittala, M., Kynkäänniemi, T., Lehtinen, J., Aila, T. \& Laine, S. Guiding a diffusion model with a bad version of itself. {\em Advances In Neural Information Processing Systems}. \textbf{37} pp. 52996-53021 (2024)

\bibitem{zhang2025cfdiffusion}Zhang, T., Zhao, Z., Ren, J., Zhang, Z., Zhang, H. \& Wang, G. cfDiffusion: diffusion-based efficient generation of high quality scRNA-seq data with classifier-free guidance. {\em Briefings In Bioinformatics}. \textbf{26}, bbaf071 (2025)

\bibitem{baron2016single}Baron, M., Veres, A., Wolock, S., Faust, A., Gaujoux, R., Vetere, A., Ryu, J., Wagner, B., Shen-Orr, S., Klein, A. \& Others A single-cell transcriptomic map of the human and mouse pancreas reveals inter-and intra-cell population structure. {\em Cell Systems}. \textbf{3}, 346-360 (2016)


\bibitem{habermann2020single}Habermann, A., Gutierrez, A., Bui, L., Yahn, S., Winters, N., Calvi, C., Peter, L., Chung, M., Taylor, C., Jetter, C. \& Others Single-cell RNA sequencing reveals profibrotic roles of distinct epithelial and mesenchymal lineages in pulmonary fibrosis. {\em Science Advances}. \textbf{6}, eaba1972 (2020)

\bibitem{mizrak2019single}Mizrak, D., Levitin, H., Delgado, A., Crotet, V., Yuan, J., Chaker, Z., Silva-Vargas, V., Sims, P. \& Doetsch, F. Single-cell analysis of regional differences in adult V-SVZ neural stem cell lineages. {\em Cell Reports}. \textbf{26}, 394-406 (2019)


\bibitem{zheng2017massively}Zheng, G., Terry, J., Belgrader, P., Ryvkin, P., Bent, Z., Wilson, R., Ziraldo, S., Wheeler, T., McDermott, G., Zhu, J. \& Others Massively parallel digital transcriptional profiling of single cells. {\em Nature Communications}. \textbf{8}, 14049 (2017)


\end{thebibliography}
\end{document}